\documentclass[aps,pre,reprint,amsmath,amssymb, superscriptaddress]{revtex4-1}
\usepackage[margin=2cm]{geometry}
\usepackage{graphicx}
\usepackage{amsmath}
\usepackage{adjustbox}
\usepackage{tikz}
\usepackage[compat=1.1.0]{tikz-feynman}
\usepackage{amsmath,amsfonts,amssymb,verbatim,float}
\usepackage{ mathrsfs }
\usepackage[utf8]{inputenc}
\usepackage{feynmf}
\usepackage{url}

\usepackage{caption}
\usepackage{subcaption}
\usepackage[T1]{fontenc}
\usepackage{natbib}
\usepackage{float}
\restylefloat{table}

\usepackage{overpic}

\begin{document}

\title{ Rewiring driven evolution of quenched frustrated signed network}

\author{Benjamin Sven Ko\v{z}i\'c}
\affiliation{Division of Theoretical Physics, Ru\dj er Bo\v{s}kovi\'c Institute, Zagreb, Croatia}
	
\author{Salvatore Marco Giampaolo}
\affiliation{Division of Theoretical Physics, Ru\dj er Bo\v{s}kovi\'c Institute, Zagreb, Croatia}

\author{Vinko Zlati\'c}
\affiliation{Division of Theoretical Physics, Ru\dj er Bo\v{s}kovi\'c Institute, Zagreb, Croatia}

\date{\today}

\begin{abstract}  
A framework for studying the behavior of a classically frustrated signed network in the process of random rewiring is developed.
We describe jump probabilities for change in frustration and formulate a theoretical estimate in terms of the master equation.
Stationary thermodynamic distribution and moments are derived from the master equation and compared to numerical simulations. 
Furthermore, an exact solution of the probability distribution is provided through suitable mapping of rewiring dynamic to birth and death processes with quadratic asymptotically symmetric transition rates. 
\end{abstract}

\maketitle

\section{Introduction}

Statistical Physics of complex networks has been applied to numerous problems in different areas of science, such as systems biology~\cite{herrgaard2008consensus,guzzi2020biological}, neurobiology~\cite{andreev2021synchronization}, sociology~\cite{jusup2022social}, economics~\cite{jackson2008social,battiston2012liaisons}, and history~\cite{kerschbaumer2020power}, just to mention a few. 

 In the framework of the statistical physics approaches to modeling complex  real  systems, a particular place belongs to spin-like models, that were naturally extended from the regular lattices to networks. 
 To provide some examples, Ising models have been used to analyze social networks~\cite{liu2010influence} and their tendency to self-organize into communities~\cite{son2006random}.
Potts' models were also used for the same purpose~\cite{Reichardt2004, kumpula2007limited} as well as in intrusion detection in secured networks~\cite{pontes2021new} and modeling multi-opinion propagation~\cite{li2022modeling}  

 Among all of them, the study of frustrated  spin models~\cite{toulouse2theory,vannimenus1977theory}  in networks plays a crucial role in understanding and analyzing complex socioeconomic systems. 
Frustrated models, describing interactions between competing entities~\cite{antal2005dynamics, antal2006social} capture the inherent conflicts, constraints, and interactions that are prevalent in social and economic contexts. 
Application of frustrated models provides insights into the emergence of collective phenomena~\cite{dorogovtsev2008critical}, decision-making processes, and the dynamics of social networks. 
These models help elucidate the causes and consequences of socioeconomic imbalances~\cite{minh2020effect}, the formation of coalitions and alliances~\cite{antal2005dynamics},  the impact of policy interventions~\cite{konig2017networks}, or evolution of welfare~\cite{ye2020passive}, to name the few.

Moreover, the study of frustrated models enables the identification of critical points and phase transitions in socio-economic systems, offering valuable information for designing resilient and sustainable solutions to societal challenges. 
Ultimately, understanding frustrated models allows us to navigate the complexities of socio-economic systems, predict their behavior, and develop strategies for promoting stability, fairness, and overall societal well-being.
Examples of this application include investigations of stable and meta-stable phase-synchronization patterns in neural networks~\cite{gollo2014frustrated}, evaluating policies to obtain wanted socioeconomic outcome~\cite{konig2017networks}, defining punctuating equilibrium in financial time series~\cite{ponzi2000criticality} etc.

Furthermore, their relevance increases when frustrated systems are combined with signed graphs, in which each edge has a positive or negative sign \cite{konig1990theory}.
Frustrated models of signed networks have found intriguing applications in deciphering intricate interactions within biological systems. 
In these systems, nodes represent entities such as genes, proteins, or even organisms, while positive and negative edges signify activation and inhibition relationships respectively. 
This nuanced approach allows researchers to model the delicate balance of molecular and cellular processes more accurately, unveiling the complexities of signal transduction~\cite{vinayagam2014integrating, 
xiang2021predicting}, gene regulation~\cite{mason2009signed, rizi2021stability}, etc. 
By integrating both positive and negative interactions, signed networks provide a holistic view of dynamic biological systems, enabling the exploration of feedback loops, bistability, and emergent behaviors. 
These networks have proven particularly useful in uncovering regulatory mechanisms~\cite{mason2009signed}, disease progression~\cite{lo2020linking}, and drug targets~\cite{zhang2018prediction}, shedding light on the underlying principles governing life's fundamental processes and offering valuable insights for biomedical research and therapeutic development.

\begin{figure}[t!]
\centering
\includegraphics[width=0.5\textwidth]{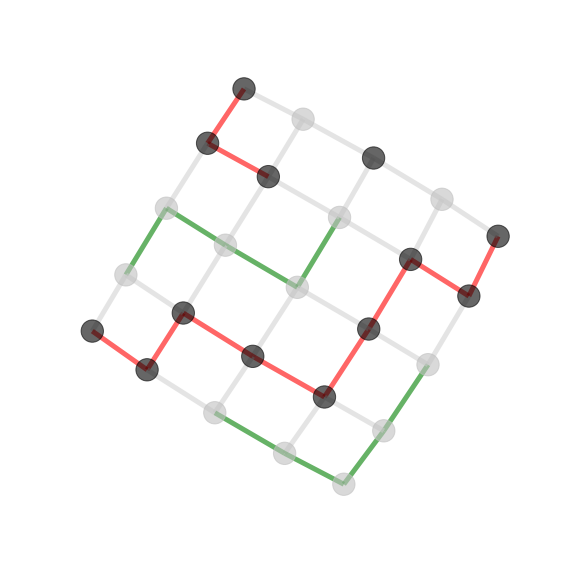}
\caption{Example of the signed lattice graph with described coloring. 
Negative frustrated links are colored red and positive frustrated links are colored green depending on the node colors.}
\end{figure}

A substantial body of literature delves into the dynamics of signed networks~\cite{lorenz2007continuous, shi2019dynamics, shi2018finite,lu2023dynamics}, primarily focusing on the dynamics of spins (signs) within these networks, with limited attention given to the evolution of network properties driven by dynamic changes. 
The pertinence of rewiring algorithms for signed networks finds its most illustrative demonstration in tools like BiRewire3~\cite{iorio2016efficient}, employed to evaluate the significance of signed network structures in biological research. 
Nevertheless, analytical outcomes for such algorithms are still outstanding, and the results hinge on intricate simulations of the systems.

Rewiring was extensively studied in the context of correlations in networks ~\cite{milo2002network,zamora2008reciprocity,zlatic2009influence}, cascading failures\cite{batool2022transition,scala2016mitigating} and especially for microcanonical sampling of networks~\cite{biely2006simulation,li2018network, vavsa2022null}.

Besides the application of frustrated models to real-world problems, their relation to different network parameters is still not solved satisfactorily. 
The problem, similar to the one analyzed in this paper, was previously addressed in \cite{campos2004frustration}. In it, the authors provide an approximate solution to the problem of evolution of ground state energy of frustrated classical Ising system with disorder modeled through the rewiring parameter of the SW model~\cite{watts1998collective}. In that paper, the authors give an approximate relation between the energy of frustration and rewiring probability $p$ of the small-world model, based on heuristic arguments. The validity of the relation was supported by simulations, except for large values of parameter $p$.  

In this paper, we comprehensively analyze how the quenched system's count of frustrated links evolves with rewiring, employing an approximate master equation. Furthermore, we demonstrate that the probability distribution of the number of frustrated bonds remains consistent with our calculated values in the thermodynamic limit.

\begin{table*}[t]
\makebox[1 \textwidth][c]{       
\resizebox{.95 \textwidth}{!}{ 

\begin{tabular}{| c | c | c | c |}
\hline
Initial configuration & Ending configuration & Change in Frustration & Approximate Probability  \\
\hline
\hline
\adjustbox{valign=c}{\begin{tikzpicture}
\tikzset{vertex/.style = {shape=circle,draw,minimum size=0.025em,fill}}
\tikzset{edge/.style = {-, > = latex}}
\node[vertex] (a) at  (0,0) [label={right:$+,-$}]{};
\node[vertex] (b) at  (0,2)[label={right:$+,-$}] {};
\node[vertex] (c) at  (2,0) [label={right:$+,-$}]{};
\node[vertex] (d) at  (2,2)[label={right:$+,-$}] {};
\draw [thick,-] (a) -- (b);
\draw [thick,-] (c) -- (d);
\end{tikzpicture}}
& \adjustbox{valign=c}{\begin{tikzpicture}
\tikzset{vertex/.style = {shape=circle,draw,minimum size=0.025em,fill}}\tikzset{edge/.style = {-, > = latex}}
\node[vertex] (a) at  (0,0) [label={right:$+,-$}]{};
\node[vertex] (b) at  (0,2)[label={right:$+,-$}] {};
\node[vertex] (c) at  (2,0) [label={right:$+,-$}]{};
\node[vertex] (d) at  (2,2)[label={right:$+,-$}] {};
\draw [dashed,-] (a) -- (d);
\draw [dashed,-] (c) -- (b);
\end{tikzpicture}}
&
 $f^\pm(t+1)=f^\pm(t)$
&---

\\
\hline
\adjustbox{valign=c}{
\begin{tikzpicture}
\tikzset{vertex/.style = {shape=circle,draw,minimum size=0.025em,fill}}
\tikzset{edge/.style = {-, > = latex}}
\node[vertex] (a) at  (0,0) [label={right:$+,-$}]{};
\node[vertex] (b) at  (0,2)[label={right:$+,-$}] {};
\node[vertex] (c) at  (2,0) [label={right:$+$}]{};
\node[vertex] (d) at  (2,2)[label={right:$-$}] {};
\draw [thick,-] (a) -- (b);
\draw [thick,-] (c) -- (d);
\end{tikzpicture}}
& \adjustbox{valign=c}{\begin{tikzpicture}
\tikzset{vertex/.style = {shape=circle,draw,minimum size=0.025em,fill}}\tikzset{edge/.style = {-, > = latex}}
\node[vertex] (a) at  (0,0) [label={right:$+,-$}]{};
\node[vertex] (b) at  (0,2)[label={right:$+,-$}] {};
\node[vertex] (c) at  (2,0) [label={right:$+$}]{};
\node[vertex] (d) at  (2,2)[label={right:$-$}] {};
\draw [dashed,-] (a) -- (d);
\draw [dashed,-] (c) -- (b);
\end{tikzpicture}} & 
 $f^\pm(t+1)=f^\pm(t)$
&---
\\
\hline
\adjustbox{valign=c}{\begin{tikzpicture}
\tikzset{vertex/.style = {shape=circle,draw,minimum size=0.025em,fill}}
\tikzset{edge/.style = {-, > = latex}}
\node[vertex] (a) at  (0,0) [label={right:$+$}]{};
\node[vertex] (b) at  (0,2)[label={right:$+$}] {};
\node[vertex] (c) at  (2,0) [label={right:$-$}]{};
\node[vertex] (d) at  (2,2)[label={right:$-$}] {};
\draw [thick,-] (a) -- (b);
\draw [thick,-] (c) -- (d);
\end{tikzpicture}}
& \adjustbox{valign=c}{\begin{tikzpicture}
\tikzset{vertex/.style = {shape=circle,draw,minimum size=0.025em,fill}}\tikzset{edge/.style = {-, > = latex}}
\node[vertex] (a) at  (0,0) [label={right:$+$}]{};
\node[vertex] (b) at  (0,2)[label={right:$+$}] {};
\node[vertex] (c) at  (2,0) [label={right:$-$}]{};
\node[vertex] (d) at  (2,2)[label={right:$-$}] {};
\draw [dashed,-] (a) -- (d);
\draw [dashed,-] (c) -- (b);
\end{tikzpicture}} &
 $f^\pm(t+1)=f^\pm(t)-1$
&
$ \cfrac{2f^{+}f^{-}}{L(L-1)}$
 \\
\hline
\adjustbox{valign=c}{\begin{tikzpicture}
\tikzset{vertex/.style = {shape=circle,draw,minimum size=0.025em,fill}}
\tikzset{edge/.style = {-, > = latex}}
\node[vertex] (a) at  (0,0) [label={right:$+$}]{};
\node[vertex] (b) at  (0,2)[label={right:$-$}] {};
\node[vertex] (c) at  (2,0) [label={right:$+$}]{};
\node[vertex] (d) at  (2,2)[label={right:$-$}] {};
\draw [thick,-] (a) -- (b);
\draw [thick,-] (c) -- (d);
\end{tikzpicture}}
& \adjustbox{valign=c}{\begin{tikzpicture}
\tikzset{vertex/.style = {shape=circle,draw,minimum size=0.025em,fill}}\tikzset{edge/.style = {-, > = latex}}
\node[vertex] (a) at  (0,0) [label={right:$+$}]{};
\node[vertex] (b) at  (0,2)[label={right:$-$}] {};
\node[vertex] (c) at  (2,0) [label={right:$+$}]{};
\node[vertex] (d) at  (2,2)[label={right:$-$}] {};
\draw [dashed,-] (a) -- (d);
\draw [dashed,-] (c) -- (b);
\end{tikzpicture}} &
 $f^\pm(t+1)=f^\pm(t)$
&---
\\
\hline
\adjustbox{valign=c}{\begin{tikzpicture}
\tikzset{vertex/.style = {shape=circle,draw,minimum size=0.025em,fill}}
\tikzset{edge/.style = {-, > = latex}}
\node[vertex] (a) at  (0,0) [label={right:$+$}]{};
\node[vertex] (b) at  (0,2)[label={right:$-$}] {};
\node[vertex] (c) at  (2,0) [label={right:$-$}]{};
\node[vertex] (d) at  (2,2)[label={right:$+$}] {};
\draw [thick,-] (a) -- (b);
\draw [thick,-] (c) -- (d);
\end{tikzpicture}}
& \adjustbox{valign=c}{\begin{tikzpicture}
\tikzset{vertex/.style = {shape=circle,draw,minimum size=0.025em,fill}}\tikzset{edge/.style = {-, > = latex}}
\node[vertex] (a) at  (0,0) [label={right:$+$}]{};
\node[vertex] (b) at  (0,2)[label={right:$-$}] {};
\node[vertex] (c) at  (2,0) [label={right:$-$}]{};
\node[vertex] (d) at  (2,2)[label={right:$+$}] {};
\draw [dashed,-] (a) -- (d);
\draw [dashed,-] (c) -- (b);
\end{tikzpicture}} &
$f^\pm(t+1)=f^\pm(t)+1$
&
$\cfrac{(L-f)(L-f-1)}{2(L(L-1))}$
\\
\hline
\hline
\end{tabular}
\caption{ Table displaying all potential network rewirings along with their associated probabilities. The initial probability constitutes a basic approximation, as it does not account for the neighboring links.} 
\label{tbl:motifs}

}}
\end{table*}

\section{Analytical results}



A frustrated classical Ising system is described in the form of a graph with $N$ nodes, with each node representing a spin that can be in one of two states (+1 or -1). 
In the terminology  used in graph theory,  this means that a set of nodes $N$ can be partitioned into two sub-sets of different colors, black (B) and white (W), where each node of spin down, $n_{i}=-1 \in B$ is colored black and spin up, $n_{j}=1 \in W$ is colored white.
Anti-ferromagnetic interaction between two spins is represented by $l$-th link, enumerating to $|L|$.
Depending on the state of two spins, we define the sign function $\sigma: l \rightarrow \{-1,0,1\}$ which tells us if the chosen link exhibits positive frustration $f^+$ i.e. link connects two nodes with positive signs, negative frustration $f^-$ i.e. a link connects two nodes with negative signs or is unfrustrated otherwise.
As such, we are considering undirected signed networks $G = (N,L,\sigma)$.

Since our use case is that all the interaction is anti-ferromagnetic, we can define \textit{frustration count} $F(G)$ as the number of frustrated links of the graph $G$ depending on the sign function $\sigma$:

\begin{equation}
F(G)=\sum_{(i,j) \in L} |\sigma(i,j)|,
\end{equation}
 where
\begin{equation}
\sigma(i,j) =
\begin{cases} 
-1 &\text{if $n_{i}=n_{j}$} = -1 \\
0 &\text{if $n_{i} \neq n_{j}$}\\
+1 &\text{if $n_{i} = n_{j}$} = \,\,\,\,1 
\end{cases}
\end{equation}

We start with the network (lattice) that has $f_0^+$ frustrations among positive spins and 
$f_0^-$ frustrations among negative spins.
For the evolution of the system, random rewiring is considered, specifically its most popular variant - the Maslov-Sneppen algorithm~\cite{maslov2002specificity} which provides a microcanonical evolution of the "small-world" network. 
The rewiring of two links is described in table \ref{tbl:motifs}. 
We chose this type of rewiring because it preserves degree distribution and can rightfully be considered as a microcanonical process, as opposed to rewiring of only one link per time step, which preserves average degrees but not exact local degrees and can therefore be considered as  equivalent to a canonical process. 


The main variable we track in the process of rewiring the network is frustration. Only a discrete number of frustration state variable $f(t)=f^{+}(t)+f^{-}(t)$ is created or destroyed in a given moment, and the simplest case is when frustration number changes by the single amount per unit time 
Here  $f^{+}(t)$ and $f^{-}(t)$ represent the number of frustrated links between positive and negative spins at time $t$. 
The frustration state of the system also depends solely on the knowledge of the most recent condition and in that case, the Markov assumption is valid, and we can formulate the evolution process in terms of the Master equation.

\subsection{Master equation for rewiring}
For integer state space differential Chapman-Kolmogorov equation~\cite{Gardiner} translates into the Master equation of the form:
\begin{eqnarray}
\delta_{t} P(f;t|f';t') &= &\sum_{k} [W(f|k;t)P(k;t|f';t')\nonumber\\
                         &   & \quad -W(k|f;t)P(f;t|f';t')]
\end{eqnarray}
Equation states that conditional probability $P(f;t|f';t')$ of state $(f,t)$ for given state $(f',t')$ at the previous time $t'<t$ is determined through summation of all other possible $k$ states and corresponding transitional probabilities $W(f|f';t)$, which can be written as:
\begin{equation}
W(f|f';t) = t^{+}(f')\delta_{f,f'+1}+t^{-}(f')\delta_{f,f'-1}
\end{equation}
In our case, using the Maslov-Sneppen rewiring algorithm,  in every single time-step the number of frustrated links can increase or decrease only by 2. Therefore we have only two possible transition probabilities, which leads to the general master equation being:

\begin{eqnarray}
\delta_{t} p(f;t) & = &t^{+}(f-2)p(f-2;t)+t^{-}(f+2)p(f+2;t)\nonumber \\
& &-[t^{+}(f)+t^{-}(f)]p(f;t).
\label{EQ: ME-Rewiring1}
\end{eqnarray}

Since Maslov-Sneppen rewiring can increase/decrease the number of frustrated links only by 2 with one of them being between positive nodes and the other between negative nodes, this means that the difference between positive and negative frustrated links is conserved in this type of rewiring i.e. $f^+-f^-=\Delta=Const$. 
 Therefore trivially we have that both $f^+$ and $f^-$ can be written as $f^\pm=(f\pm\Delta)/2$
We can now specify functions $t^{\pm}(f)$ which describe the probability of change in frustration by consulting the table \ref{tbl:motifs}.

In the following, we will mainly use the probabilities of choices that are changing the number of frustrated links since they are needed to specify the transition probabilities, while other choices.
The probability that frustration decreases means that we choose 1 out of all positive links, and 1 out of all negative links, divided by the number of all combinations of pairs of links. The difference in positive and negative frustration $\Delta$ is a key parameter that distinguishes an ensemble of graphs.

\begin{eqnarray}
t^{-}(f) &=& \frac{\binom{f^{+}}{1}\binom{f^{-}}{1}}{\binom{L}{2}} = \frac{f^{2}-\Delta^{2}}{2L(L-1)}\\
t^{+}(f) &=& \frac{\binom{L-f^{+}-f^{-}}{2}}{2\binom{L}{2}} = \frac{f^{2} + (1-2L)f + L(L-1)}{2L(L-1)}\nonumber 
\end{eqnarray}
The probability of an increase in frustration via dual rewiring is the number of all unfrustrated pairs divided by the number of all pairs.
These approximated probabilities were confirmed by running simulations of choosing two links, rewiring them, and comparing the increase or decrease in frustration. Such an approximation is good enough to compute the evolution of frustrated links subject to the Maslov-Sneppen rewiring. In table \ref{tbl:motifs}, we provide additional more precise approximations that might be of use for extremely small network systems, but that we did not use in the rest of this study.
\subsection{Stationary solution}

If the limit of a large network, the solution of (\ref{EQ: ME-Rewiring1}) is going to converge to simulations. We can integrate this solution directly by numerical integration, using a choice of boundary probabilities $p(f<\Delta;L,\Delta,t)=0$ and $p(f>L;L,\Delta,t)=0$
with initial condition $p(f;L,\Delta,t=0)=\delta(f_0)$. We can also use thermodynamical limit ($N\rightarrow\infty)$ to approximate the difference between probability distributions at time $t+1$ and $t$ as: $p(f,t+1)-p(f,t)\approx\dot{p}(f,t)$. Using generating functions $G(z,t)=\sum_{f=\Delta}^L z^fp(f,L,\Delta,t)$, one can see that $\dot{G}(z,t)=\sum_{f=\Delta}^L z^f\dot{p}(f,L,\Delta,t)$. The usual properties of generating function hold: 
\begin{flalign}
    &G(0,t)=0\nonumber\\
    &G(1,t)=1\nonumber\\
    &p(f,L,\Delta,t)=\frac{\partial^fG(z,t)}{\partial z^f}|_{z=0}\nonumber\\
    &\mathbf{E}(f(t))=\frac{\partial G}{\partial z}|_{z=1}.
\end{flalign}

\noindent First, we calculate the simpler case using stationarity condition $\lim_{t\rightarrow\infty}\dot{p}(f,t)=0$ to obtain the solution in the form of a hypergeometric function.

\begin{eqnarray}
 G(z;L,\Delta) &=&\frac{z^\Delta \, _2F_1\left(\frac{\Delta-L}{2},\frac{(\Delta-L+1)}{2};\Delta+1;z^2\right)}{\, _2F_1\left(\frac{\Delta-L}{2},\frac{(\Delta-L+1)}{2};\Delta+1;1\right)} \nonumber \\
    &=&\frac{\sqrt{\pi } 2^{-\Delta-L} \Gamma (d+L+1) z^{\Delta}}{\Gamma (\Delta+1) \Gamma
   \left(L+\frac{1}{2}\right)}\times\\& &\times\sum_{n=0}^{\infty} \frac{\left(\frac{\Delta-L}{2}\right)_n \, \left(\frac{(\Delta-L+1)}{2}\right)_n}{(\Delta+1)_n} \frac{(z^2)^n}{n!}\nonumber
\end{eqnarray}
In the case in which $\Delta\rightarrow 0$
the generating function $\lim_{\Delta \rightarrow 0}{G(z;L,\Delta)} = G(z;L)$ will depend only on the parameter $L$:
\begin{equation}
G(z;L)=\frac{_2F_1(\frac{1-L}{2}, -\frac{L}{2}, 1, z^2)}{_2F_1(\frac{1-L}{2}, -\frac{L}{2}, 1, 1)}
\end{equation}
 From this expression, we can obtain the probability distribution using an inverse Z-transform. 

For $f$ even and in the interval $f\in[0,L]$, we obtain that the stationary probability is equal to  
\begin{equation}
  p(f;L,\Delta=0) =  \frac{\sqrt{\pi } 2^{-f-L} \Gamma (L+1) \Gamma (f-L)}{\Gamma \left(\frac{f}{2}+1\right)^2 \Gamma (-L)
   \Gamma \left(L+\frac{1}{2}\right)},
\label{probability}
\end{equation}
while vanishes in all the other cases.
The expected value of this stationary distribution for $L>0$ is simply:
\begin{equation}
    \mathbf{E}\left[f;L,\Delta=0,t=\infty \right] = \frac{L(L-1)}{2L-1}
\label{eq::exp}
\end{equation}



\begin{figure}[t]
\centering
\includegraphics[width=0.9\linewidth]{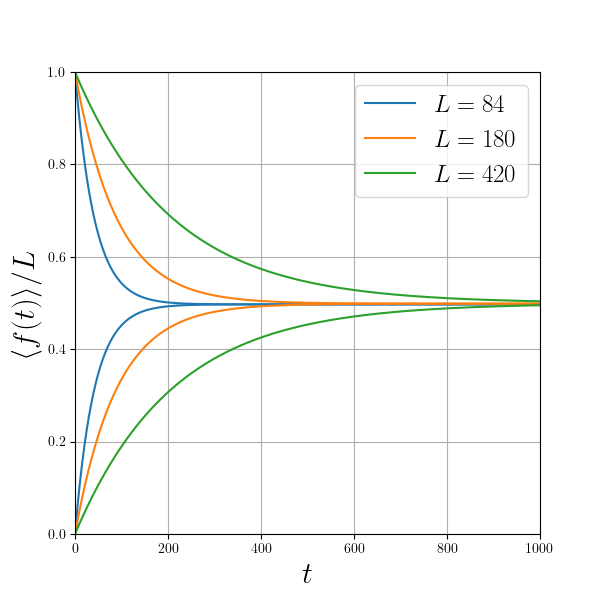}
\caption{Evolution of mean value of frustration through the process of rewiring from boundary values of initial frustration  $f_{\text{min}}=0$ and $f_{\text{max}}=L$ for $L = 84, 180, 420$, with $\Delta = 0$.}
\label{fig:evolution}
\end{figure}


When this discrete stationary probability distribution $p(f;L,\Delta)$ is evaluated for large enough parameter $L$ and with $\Delta=0$, it tends to normal distribution.\\

\begin{figure}[t]
\centering
\begin{subfigure}{.5\textwidth}
  \centering
  \includegraphics[width=0.9\linewidth]{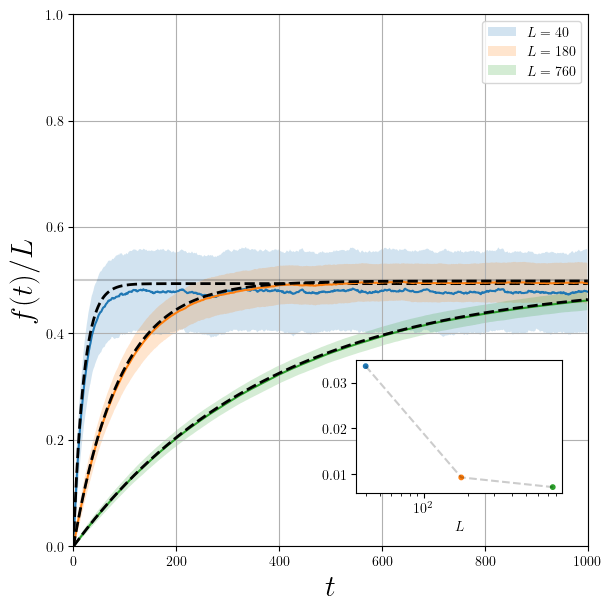}
\end{subfigure}%
\newline
\begin{subfigure}{.5\textwidth}
  \centering
  \includegraphics[width=0.9\linewidth]{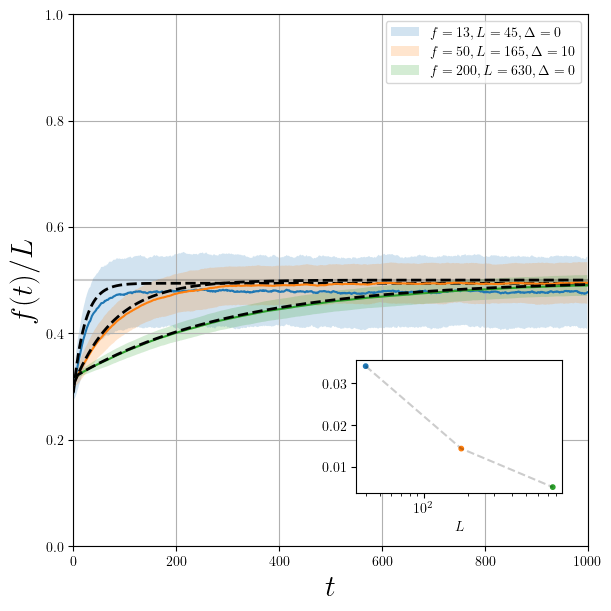}

\end{subfigure}

\caption{(a) Rewiring process for a square lattice starting from the lowest possible frustration $f=0$ (b) same rewiring process for triangular lattice starting from $f>0$. The inset axis shows the mean absolute error decreasing for larger system size $L$.}
\label{Convergence}
\end{figure}

\subsection{Evolution of moments}

Though determining the precise distribution $p(f,t)$ is intricate, calculating the expected value $f(t)$ from the differential equation in the generating function is straightforward.
The general solution takes the form:

\begin{eqnarray}
\langle f(t;f_{0},L,\Delta)\rangle &=& \left(f_{0} - \frac{\Delta^2 + L(L-1)}{2L-1}\right)e^{-\frac{(2L-1)}{L(L-1)}t}\nonumber\\& &+\frac{\Delta^2+L(L-1)}{2L-1}
\label{eq:result}
\end{eqnarray}
The value of frustration depends on the parameters $f_{0},L,\Delta$. 
The stationary state in the limit of a large network is described as $\langle f\rangle\sim\frac{L}{2}+\Delta^2\frac{1}{L}$. In models in which $\Delta\sim \mathcal{O}(L)$ the deviation from the trivial solution will be governed solely by the difference between positively and negatively frustrated bonds.

Although the parameter $\Delta$ influences the final expected value of the evolution, it does not influence the rate with which the system achieves its thermodynamic state. 
However, for $\Delta=0$, we have two initial boundary values: $f_{\text{min}}$ and $f_{\text{max}}$, both approaching the same expected value (\ref{eq::exp}) as $t \rightarrow \infty$ (as shown in figure \ref{fig:evolution}). 

\subsection{Probability distribution evolution}

A stochastic birth and death process can be described using the Chapman-Kolmogorov equations:
\begin{equation}
\frac{\mathrm{d} p_{n}(t)}{\mathrm{d} t} = \lambda_{n-1}p_{n-1}(t) + \mu_{n+1}p_{n+1}(t)-(\lambda_{n}+\mu_{n})p_{n}(t) 
\end{equation}
 where $\lambda_{n}$ and the $\mu_{n}$ are respectively the birth and death rates. 
In the analysis of our rewiring process, we can use the same equation by defining 

\begin{eqnarray}
\lambda_{f^{\pm}} = \frac{\binom{L-f^{+}-f^{-}}{2}}{2\binom{L}{2}} &\rightarrow& \lambda_{f}=\frac{f^{2} + (1-2L)f + L(L-1)}{2L(L-1)} \nonumber \\
\mu_{f^{\pm}} = \frac{\binom{f^{+}}{1}\binom{f^{-}}{1}}{\binom{L}{2}} &\rightarrow& \mu_{f}=\frac{f^{2}-\Delta^{2}}{2L(L-1)}
\end{eqnarray}

Since the master equation which governs the process of rewiring always adds or subtracts 2 frustrated links, we introduce variable $\varphi$, such that  $2\varphi=f$. This leads to the new equation
\begin{equation}
\frac{\mathrm{d} p_{\varphi}(t)}{\mathrm{d} t} = \lambda_{\varphi-1}p_{\varphi-1}(t) + \mu_{\varphi+1}p_{\varphi+1}(t)-(\lambda_{\varphi}+\mu_{\varphi})p_{\varphi}(t). 
\end{equation}
In which 
\begin{align}
\lambda_{\varphi}&=\frac{4\varphi^{2} - 2(2L-1)\varphi + L(L-1)}{2L(L-1)},\nonumber\\
\mu_{\varphi}&=\frac{4\varphi^{2}-\Delta^{2}}{2L(L-1)}
\label{EQ: LambdaMu1}
\end{align}
Noticing that the smallest number of frustrated links possible in the system is equal to $\Delta$, we re-parametrize it so that $\delta=\Delta/2$, and introduce variable $\phi=\varphi-\delta$, which leads to parameters:
\begin{align}
\lambda_{\phi}&=\frac{4(\phi+\delta)^{2} - 2(2L-1)(\phi+\delta) + L(L-1)}{2L(L-1)},\nonumber\\
\mu_{\varphi}&=\frac{4(\phi^{2}+2\phi\delta)}{2L(L-1)}
\label{EQ: LambdaMu2}
\end{align}

This particular set of parameters corresponds to the birth and date process previously analyzed in the work of Roehner and Valent~\cite{Roehner-Valent}  which solved  the general solution for the transition rates:
\begin{equation}
\lambda_{n} = \alpha (n^2 + bn + c)=\alpha(n-\nu_+)(n-\nu_-), \mu_{n} = \alpha(n^2 + \tilde{b}n)
\label{gen-eq}
\end{equation}

Our process then has parameters defined as:
\begin{align}
\alpha &= \frac{2}{L(L-1)},b=\frac{(4\delta-2L+1)}{2},\nonumber\\ c &=\frac{4\delta^2-(4L+2)\delta+L(L-1)}{4}, \tilde{b}=2\delta    
\end{align}

This redefinition of transition rates corresponds to~\cite{Roehner-Valent} analysis of 
 finite processes where the population can only evolve between the levels $0$ and $N=L/2$. If we take a large limit so that $ L^2 \gg L $\\

The general formula derived is:
\begin{align}
    G(z,t) =& \sum_{l=0}^{\left [ a \right ]} H_{l} y_{l}^{(1)}(z)e^{-l(2a-l)\alpha t}\nonumber\\ &+\int_{0}^{\infty}h_{n_{0}}(u)y_{u}^{(2)}(z)e^{-(u^2-a^2)\alpha t}du
\end{align}
where $\left [ a \right ] =$ the integer part of $a$ and $H_{l}$ being:

\begin{eqnarray}
    H_{l} &=& (-1)^{l+1}\frac{2(l-a)\sin{2\pi(l-a)}}{\pi \Gamma{(1+\tilde{b})}}\nonumber\\
    &&\times\Gamma{(l-\nu_{+})}\Gamma{(l-\nu_{-})}\Gamma{(2a-l-\nu_{+})}\Gamma{(2a-l-\nu_{-})}\nonumber\\
    &&\times \sum_{r=0}^{\text{inf}(n_{0},l)}\frac{\binom{n_{0}}{r}}{(l-r)!}\frac{\Gamma{(l+r-2a)}}{\Gamma{(r-\nu_{+})}\Gamma{(r-\nu_{-})}}
\end{eqnarray}
Details of the computation are provided in the appendix.

The second part of $G(x,t)$, $h_{n_{0}}(u)$ vanishes if $n_{0}\leq N$.
For transition rates defined with our transformation, we obtain time-dependent probability distribution, which for $\delta=0$ is easy to write as:
\begin{flalign}
   p(\phi,t) &= -\sum_{l=0}^{L/2} \mathcal{F}(l,\phi,L) e^{\frac{t (2 l-L) (2 l+L+1)}{2 (L-1) L}} 
\end{flalign}
Where function $\mathcal{F}(l,\phi,L)=A(l,\phi,L)B(l,L)C(l,\phi,L)$ is split into three parts for simplicity:
\begin{flalign}
       A(l,\phi,L)&= \frac{ \Gamma (L+2) \Gamma
   \left(\phi-l-\frac{L}{2}-\frac{1}{2}\right)}{\sqrt{\pi } (L+1) \Gamma (\phi+1) \Gamma
   \left(-l+\frac{L}{2}+1\right)}\nonumber\\
       B(l,L)&= (-1)^l \cos \left(\frac{\pi  L}{2}\right)(4 l+1) 2^{-L-1}\\
       C(l,\phi,L)&=\!\, _3F_2\left(\!-\phi,\!l+\frac{1}{2},\!l+1;1,\!-\phi+l+\frac{L}{2}+\frac{3}{2};\!1\right).\nonumber
\end{flalign}
Where $ _3F_2\left(\!-\phi,\!l+\frac{1}{2},\!l+1;1,\!-\phi+l+\frac{L}{2}+\frac{3}{2};\!1\right)$  is a hypergeometric function.

\section{Validation}

To test the analytical solutions, numerical simulations were executed exploiting the Python Networkx package. 
Our main objective is to validate that our estimated analytical model aligns with the outcomes of simulations for two scenarios: the transition from a state with 1) minimal frustration and 2) maximal frustration. 
In the second scenario, it's important to note that this isn't the stationary state of maximum conceivable frustration, but rather a state where a change in frustration value is plausible, i.e. the most likely outcome.

\subsection{Minimum and maximum Frustration}

First, we randomly assign spin values to the nodes of a graph of size $L$ to determine the $\Delta$ parameter. 
We then optimize the graph to minimize frustration and obtain the system configuration. 
Starting from the configuration set by parameters $(L,f_{0},\Delta_{0})$, we proceed with the rewiring process for $t=3000$ steps. 
At each step $t$, we calculate the system's frustration, repeating this process for $n=1000$ iterations. 
The mean value and standard deviation are obtained by averaging $n$ frustration values: $\langle f(t)\rangle=\frac{1}{n}\sum_{i=0}^{n}f_{i}(t)$ for each time step $t$.

In the case of a square grid, achieving a minimum frustration value of $f=0$ is always possible. 
However, for a triangular grid, the minimum frustration is not zero.
We can observe that the theoretical solution approximates the rewiring process effectively. 
Accuracy improves with larger system sizes, evident from the reduced mean absolute error between the analytical function and the average frustration value. 
The systems' frustration tends towards the expected theoretical value.

\begin{figure}[t]
\centering
\begin{subfigure}{.5\textwidth}
  \centering
  \includegraphics[width=0.75\linewidth]{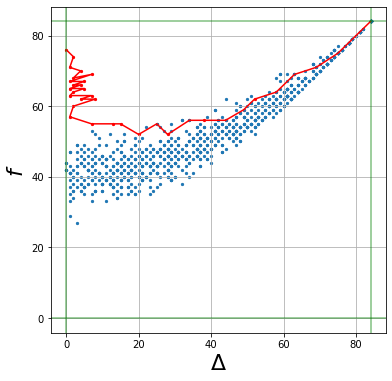}

\end{subfigure}%
\newline
\begin{subfigure}{.5\textwidth}
  \centering
  \includegraphics[width=.75\linewidth]{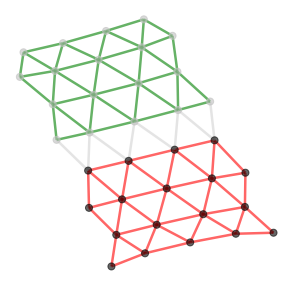}

\end{subfigure}
\caption{(a) A plot of ($f,\Delta$) configurations. Blue points represent the randomly initialized configuration sample of 1000 graphs, red trajectory shows the simulated annealing process from starting point $(84,84)$ to an optimum solution (b) Solution to optimization of a $L=84$ triangular graph with $f=76,\Delta=0$}
\label{fig:test}
\end{figure}

\begin{figure}[t]
\begin{subfigure}{.5\textwidth}
  \centering
  \includegraphics[width=.9\linewidth]{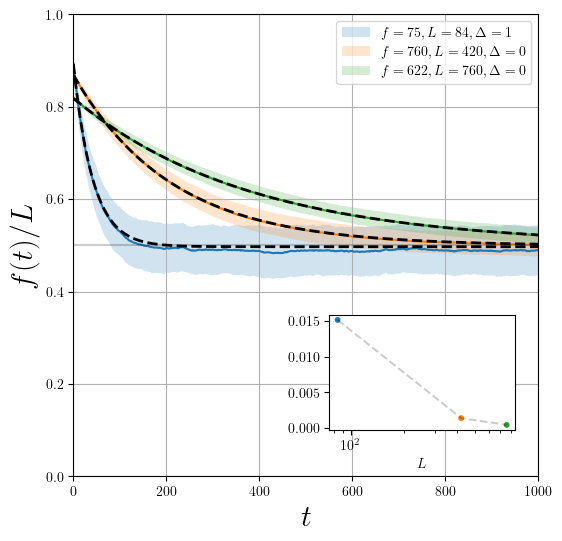}
\end{subfigure}%
\newline
\begin{subfigure}{.5\textwidth}
  \centering
  \includegraphics[width=.9\linewidth]{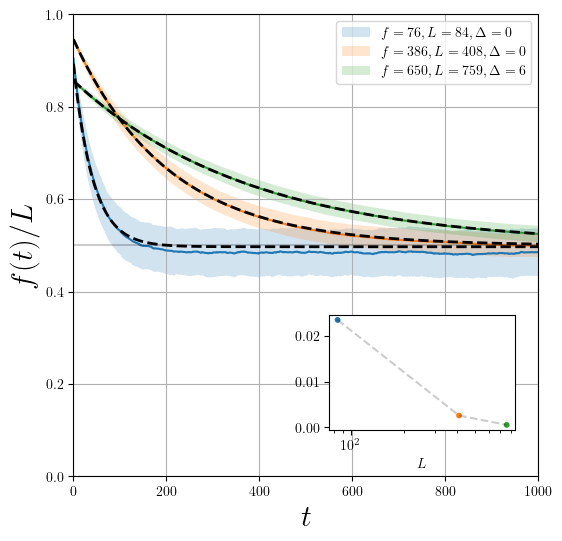}
\end{subfigure}
\caption{Rewiring process starting from the approximately maximum obtainable frustration and lowest $\Delta$ value for square and triangular lattice.}
\label{fig:max}
\end{figure}
Clearly, a system that has $f=L, \Delta=L$ won't exhibit the behavior of maximum possible decrease in frustration by process of rewiring noted in figure \ref{fig:evolution}.

For a system to be in the state of maximum frustration and allow for this type of evolution, two conditions need to be satisfied, namely that $f_{0} = L$ and that the difference between positive and negative frustrated links $\Delta = 0$. 
To achieve that, a simulated annealing algorithm was used with a Euclidean distance cost function that targets maximum frustration and minimum $\Delta$.
\begin{equation}
    \mathbf{y} = (f,\Delta) \rightarrow d(\hat{\mathbf{y}},\mathbf{y}) = \sqrt{(f-\hat{f})^2+(\Delta - \hat{\Delta})^2} 
\end{equation}
Where $\hat{\mathbf{y}} = (L,0)$ is the target value of the system state.
This condition would ideally mean that the graph has two separate parts with $f^{+}+f^{-} = L$. 
Not all graph configurations can satisfy these two conditions, and for them, the next optimum solution is chosen, an example is shown in figure \ref{fig:test}.

The value of average frustration decreases as predicted in figure \ref{fig:max}. The ratio $f/L_{\square} = 0.89, 0.85, 0.81$ and $f/L_{\triangle} = 0.90, 0.95, 0.86$ for square lattice and triangular lattice respectively denote the initial starting points. 
Because of the stochastic behavior of simulated annealing and the implication of the size of the system, in practice, these ratios are slightly lower for larger systems.


\begin{figure}[t]
\centering
\includegraphics[width=0.48\textwidth]{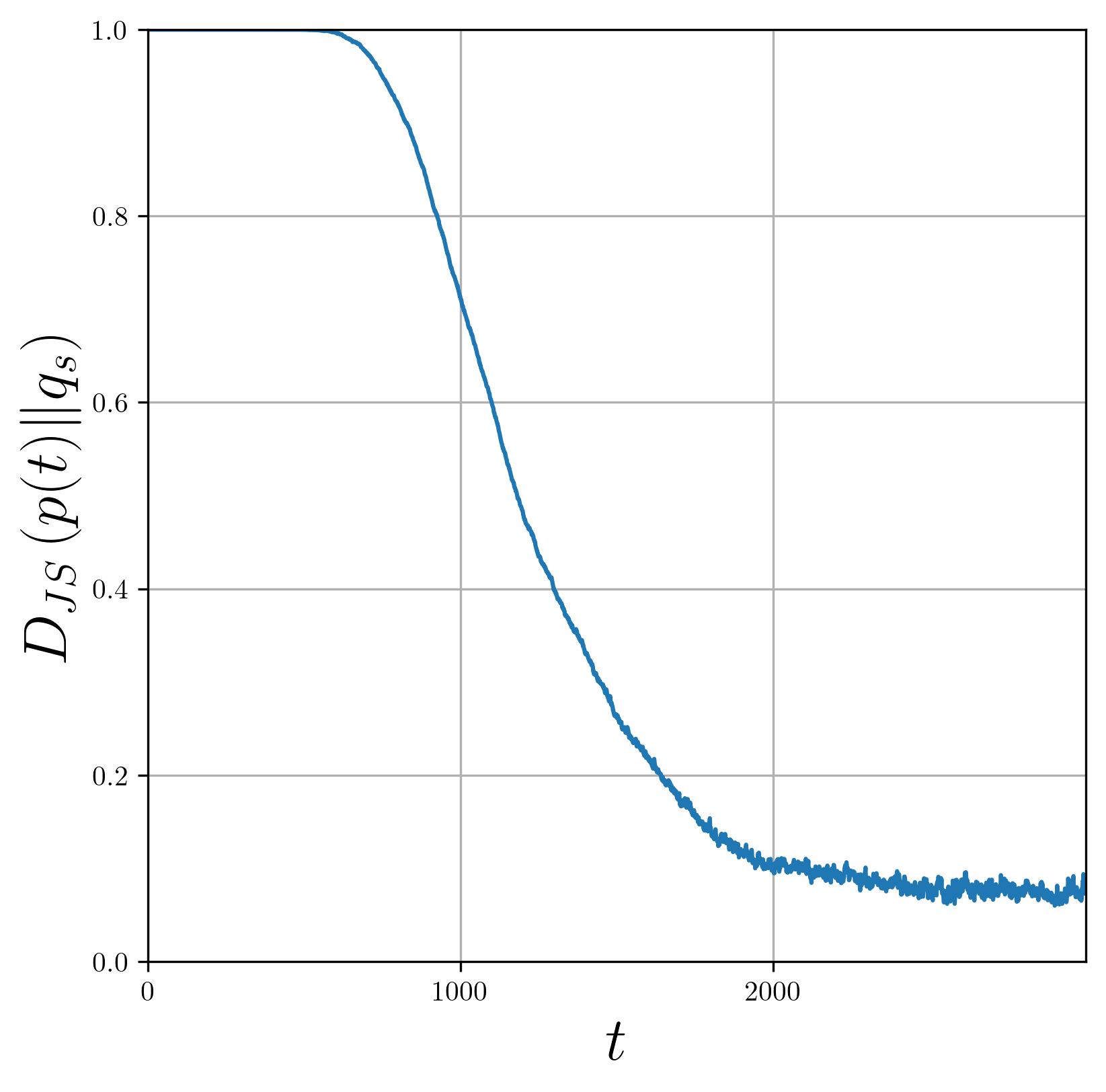}
\caption{Jensen - Shannon Divergence between asymptotic analytical $p(f)$ distribution and simulated average distribution $q(f,t) $.}
\label{JSD}
\end{figure}

\subsection{Probability distributions}

To calculate the probability distribution of a system, we have several options, one of which is to assume that the process is ergodic after a certain period of rewiring, arriving at the stationary state. We count the values of frustration for each of the $n$ systems, therefore obtaining the probability distribution, and take the average over the stationary period $\Delta t = t_{2}-t_{1}$.\\
\noindent This is shown on the right plots of figures \ref{max} and \ref{min} (in appendix) for the square and triangular lattice of similar size for increase and decrease of frustration by the process of rewiring, respectively.\\

Red lines in figures \ref{max} and \ref{min} (in appendix) denote the means and the standard deviation calculated from the analytic distribution which matches the simulated values nicely, suggesting that, although the theoretical model is only approximate, the extra information contained in the probability distribution describes the process well for larger systems.
 
\begin{figure}[t]
\includegraphics[width=0.48\textwidth]{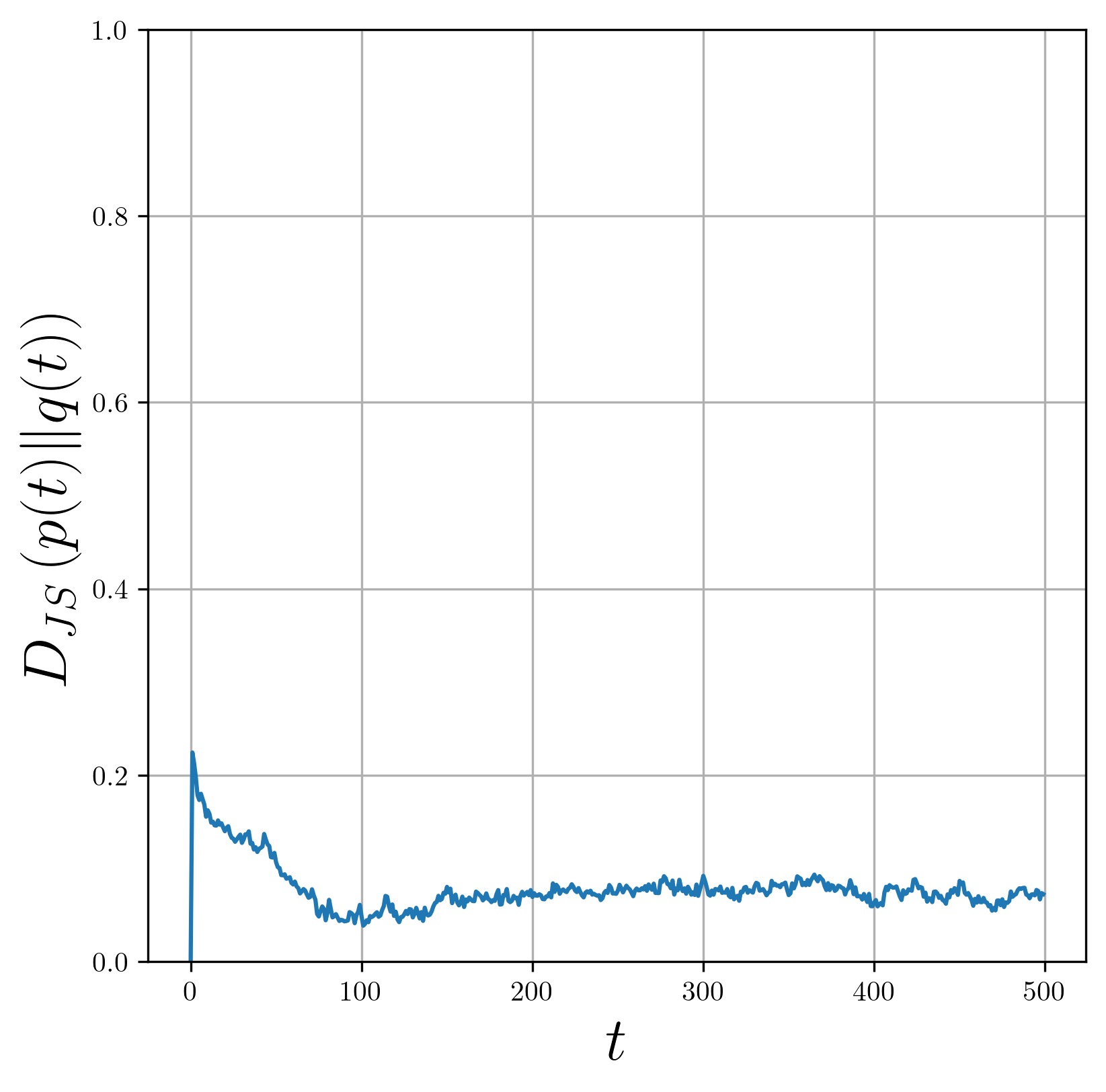}
  \caption{Jensen-Shannon divergence of the simulated and analytically evaluated probability distribution at time $t$. Initial jump is related to the initial difference between evolutions mainly governed by the difference between discrete time steps of simulated probability distributions and the continuous time steps of the master equation solution.}
  \label{Fig:ProbEvo}
\end{figure}
A second way to compare the simulated probabilities with analytical results, depending on the time in the rewiring process, is to sample all $n$ evolutions in $k$  groups and count the occurring number of all possible frustrations ($f=0...,L$) in each of the $k$ systems for every time step $t$. In simulations we mainly used $k=100$. We can then calculate the average probability distribution $\langle p(f,t) \rangle$ and its standard deviation.

To measure the similarity between the theoretical stationary distribution and the simulated one, we use Jensen-Shannon divergence:

\begin{equation}
    D_{JS}(p || q) = \frac{1}{2}\text{D}(p || \frac{1}{2}(p+q)) + \frac{1}{2}\text{D}(q || \frac{1}{2}(p+q))
\end{equation}
Where $\text{D}(p || q)$ is the Kullback-Leibler divergence:
\begin{equation}
    \text{D}(p || q) =\sum_{x}p(x) \log \left ( \frac{p(x)}{q(x)} \right ) 
\end{equation}
\begin{figure}[t]
\includegraphics[width=0.5\textwidth]{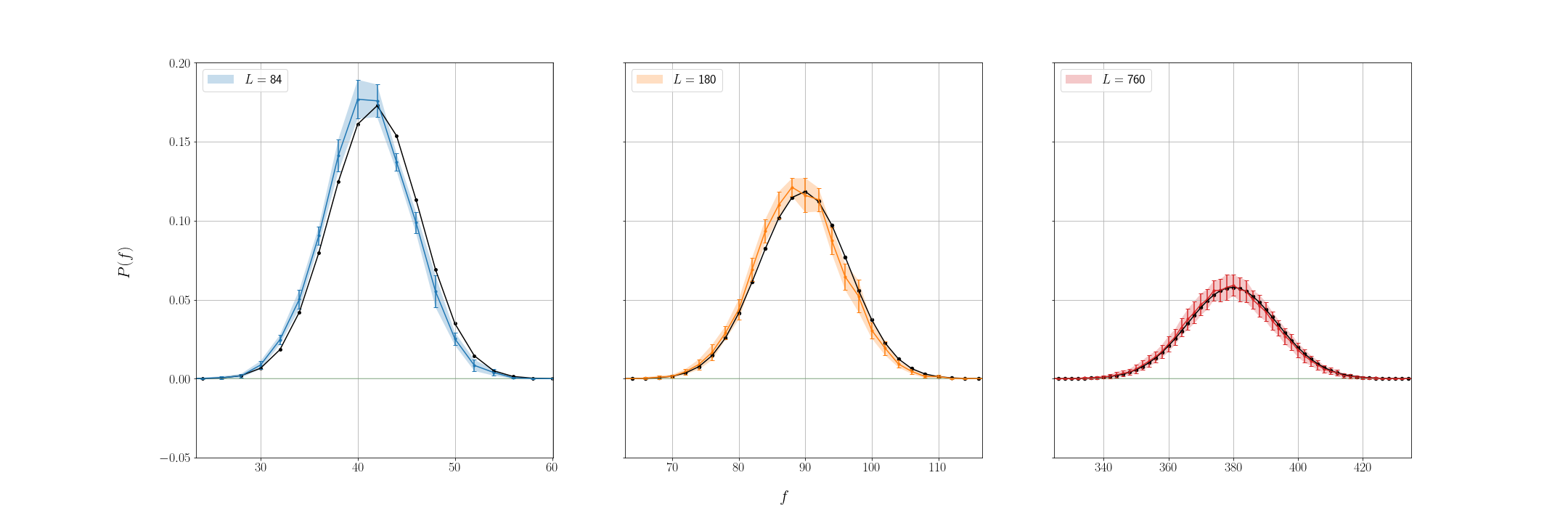}
  \caption{Probability of a system assuming frustration value $f$ in the stationary regime depending on the size of the system $L$. Simulation is represented in color, while the analytic solution is black.}
  \label{fig:key}
\end{figure}

In figure \ref{JSD} we see that this metric measure decreases from the value of 1 representing the difference between the certain probability of system in state $f_{0}$ and the stationary distribution to a smaller nonzero value, implying that the distributions become similar as the rewiring proceeds.

In order to demonstrate that our analytical approach can follow the evolution of the whole probability distribution, we present figure \ref{Fig:ProbEvo}. The initial steep increase of the Jensen-Shannon divergence is related to the discrete nature of time steps for the simulations, which gets smoothed out during the evolution as the number of time steps becomes larger.   

Here an important Remarque is in order. Assuming that the stationary probability $p(f;L,\Delta)$  is the probability that a given system has frustration $f$, these values do not depend on the time $t$ of the rewiring process.
If we sample $n=1000$ systems with random configurations of $\Delta$ we can also obtain the probability distribution of frustration for a system of size $L$. It is clear that the stationary probability distribution obtained by the process of rewiring will coincide with this probability distribution, as is shown in figure \ref{fig:key}.

\section{Conclusion}

In this paper, we presented an estimated solution for the energy of a frustrated Ising antiferromagnetic model based on the disorder level in the system. Though earlier studies offered some approximate solutions, our work introduces a more accurate solution. Additionally, we present supporting evidence that demonstrates the precise agreement between our solution and an exact solution in the thermodynamic limit. 

From a purely physical perspective, we have investigated the quenched spins, in a fast-changing network as opposed to another limit of slowly changing network and fast flipping spins studied in ~\cite{campos2004frustration}. While the quenched frustration energy changes with $\sim e^{-\frac{(2L-1)}{L(L-1)}t}$, the previously studied ground state energy per spin changes with $t$. An interpolating behavior between these 2 regimes that were not previously studied is of particular interest. We envision analyzing such a cross-over regime in the future. 

Computed evolution of the frustration subjected to random rewiring provides other researchers with important information that can be used in a number of different applications. For example, provided solutions could be used in testing algorithms in extremely large networks. Furthermore, although we focused our attention on frustrated links in signed networks, a very similar analysis can be used in networks with differently colored edges, like for example in multiplex networks~\cite{gomez2013diffusion,battiston2014structural,bianconi2013statistical,lu2023dynamics,kenett2015networks}. We believe that it can be especially useful in the application of frustrated models in large changing network systems, such as social, biological, or technological networks.

\begin{acknowledgments}
The authors acknowledge support from the Croatian Science Foundation (HrZZ) Projects No. IP–2019–4–3321.
S.M.G. and V.Z. also thanks the QuantiXLie Center of Excellence, a project co–financed by the Croatian Government and European Union through the European Regional Development Fund – the Competitiveness and Cohesion (Grant KK.01.1.1.01.0004). 
\end{acknowledgments}


\bibliography{mainArxiv}

\onecolumngrid

\section*{Appendix}

\subsection{Time-dependent Generating function}

Here we briefly outline the procedure for obtaining the general solution for transition rates describing the so called "quadratic asymptotically symmetric processes" \cite{Roehner-Valent}\cite{letessier1984generating}.

\begin{flalign}
\lambda_{n} &= \alpha (n^2 + bn + c) = \alpha(n-\nu_{+})(n-\nu_{-})\\
\mu_{n} &= \alpha(n^2 + \tilde{b}n)
\end{flalign}

\noindent Generating function is the solution of the hypergeometric equation:
\begin{gather}
    \frac{\partial G}{\partial t}(z,t) = \alpha(1-z)\left[z(1-z)\frac{\partial^{2}}{\partial {z}^{2}}+(1+\tilde{b}-(1+b)z)\frac{\partial}{\partial z}-c\right]G(z,t)\\
    G(z,0) = z^{n_{0}}, n_{0}\in \mathbb{N}, z\in\left[0,R\right], R<1\\
    G(1,t) = 1, t\geq 0
\end{gather}
Which is solved using the eigenvectors $y$ of the associated hypergeometric operator $\mathcal{L}_z$, defined through $ \frac{\partial G}{\partial t}(z,t) =\mathcal{L}_zG$.
The general equation for generating function is:
\begin{equation}
    G(z,t) = \sum_{l=0}^{\left [ a \right ]} H_{l} y_{l}^{(1)}(z)e^{-l(2a-l)\alpha t}+\int_{0}^{\infty}h_{n_{0}}(u)y_{u}^{(2)}(z)e^{-(u^2-a^2)\alpha t}du
\end{equation}
where $\left [ a \right ] =$ integer part of $a$ and $H_{l}$ being:
\begin{gather*}
    H_{l} = (-1)^{l+1}\frac{2(l-a)\sin{2\pi(l-a)}}{\pi \Gamma{(1+\tilde{b})}}\Gamma{(l-\nu_{+})}\Gamma{(l-\nu_{-})}\Gamma{(2a-l-\nu_{+})}\Gamma{(2a-l-\nu_{-})}\\
    \times \sum_{r=0}^{\text{inf}(n_{0},l)}\frac{\binom{n_{0}}{r}}{(l-r)!}\frac{\Gamma{(l+r-2a)}}{\Gamma{(r-\nu_{+})}\Gamma{(r-\nu_{-})}}
\end{gather*}
with:
\begin{eqnarray}
    \nu_{\pm} = -\frac{b}{2} \pm \partial, a = \frac{1+\tilde{b}-b}{2}=\frac{1+\tilde{b}+\nu_{+}+\nu_{-}}{2}, \partial = \sqrt{\left (\frac{b}{2}\right)^{2} - c^2}
\end{eqnarray}
As already specified, mapping our rewiring problem to this class requires substitution $2(\phi+\delta)=f$: 

\begin{align}
\alpha &= \frac{2}{L(L-1)}, b=\frac{(4\delta-2L+1)}{2}, c=\frac{4\delta^2-(4L+2)\delta+L(L-1)}{4}, \tilde{b}=2\delta    
\end{align}
We can showcase that by taking the limit $\delta\rightarrow 0$,
\begin{align}
\alpha &= \frac{2}{L(L-1)}, b=\frac{1-2L}{2}, c=\frac{L(L-1)}{4}   
\end{align}
produces the following generating function $G(z,t)$:
\begin{equation}
    G(z,t) = -\frac{\Gamma \left (\frac{L}{2}+1 \right )}{2 \Gamma \left (\frac{1}{2} - \frac{L}{2} \right )}\sum_{l=0}^{L/2}\frac{(-1)^l (4 l+1)\Gamma
   \left(-l-\frac{L}{2}-\frac{1}{2}\right)}{\Gamma \left(-l+\frac{L}{2}+1\right)}  \, _2F_1\left(l+\frac{1}{2},l+1;1;z\right) (1-z)^{\frac{1}{2} (2 l+L+1)} e^{\frac{t (2 l-L) (2 l+L+1)}{2 (L-1) L}} 
\end{equation}
and furthermore, for $t \rightarrow \infty, l \rightarrow L/2 $ we have:
\begin{gather}
     G(z) = \frac{2^{-L} \cos (\pi  L) \Gamma \left(\frac{1}{2}-L\right) \Gamma (L+1)}{\sqrt{\pi }} \, _2F_1\left(\frac{1-L}{2},-\frac{L}{2};1;z\right)
\end{gather}
Which is precisely the stationary solution.

\subsection{evolutions from minimally and maximally frustrated systems}





\begin{figure}[h]
\centering
\includegraphics[width=0.8\textwidth]{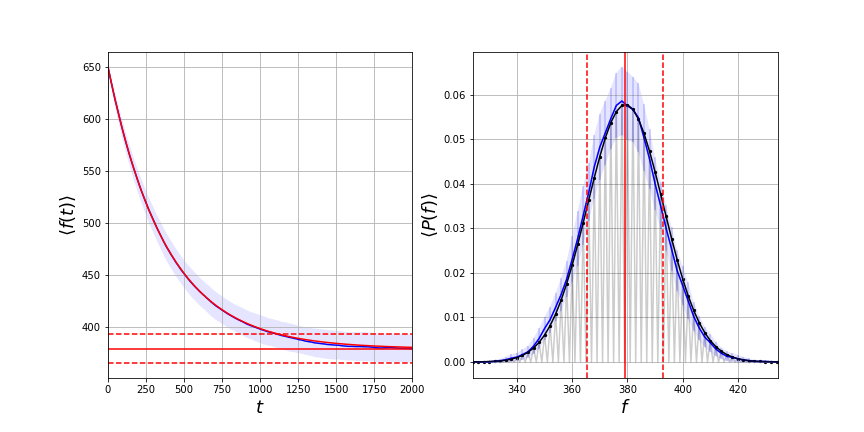}
\caption{(a) Rewiring process of \textbf{triangular lattice} for $f_{0}=650,L=759,\Delta=6$ (b) Probability distribution function averaged over time period $t_{1}=2000, t_{2}=3000$ with theoretical stationary probability distribution shown in black. Red full and slashed lines represent the mean value and standard deviation calculated from the theoretical solution. Gray zig-zag lines represent the true probability distribution as system can only obtain even values of frustration, while $p(f_{odd})=0$. For clarity, the comparison between distributions is presented in the even-envelope of distributions.}
\label{max}
\end{figure}

\begin{figure}[h]
\centering
\includegraphics[width=0.8\textwidth]{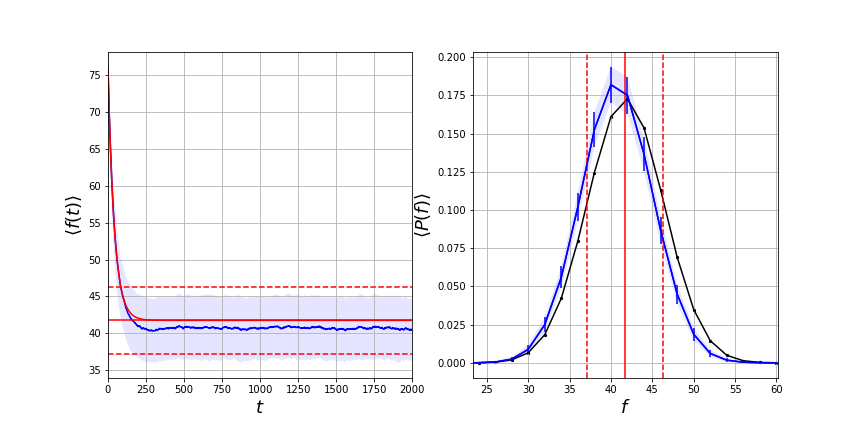}
\caption{(a) Rewiring process of square lattice for $f_{0}=0,L=760,\Delta=0$ (b) Probability distribution function averaged over time period $t_{1}=2000, t_{2}=3000$ with theoretical stationary probability distribution shown in black. Red full and slashed lines represent the mean value and standard deviation calculated from the theoretical solution.}
\label{min}
\end{figure}

\end{document}